# Photoinduced topological phase transitions in Kitaev-Heisenberg honeycomb ferromagnets with the Dzyaloshinskii-Moriya interaction

Zhengguo Tang, Heng Zhu, Hongchao Shi, Bing Tang[*]

*Department of Physics, Jishou University, Jishou 416000, China*

We theoretically study topological properties of Floquet magnon in a laser-irradiated Kitaev-Heisenberg honeycomb ferromagnet with the Dzyaloshinskii-Moriya interaction by means of the Floquet-Bloch theory. It is found that the Kitaev-Heisenberg ferromagnet can reveal two topological phases with different Chern numbers when it is irradiated by a circular-polarized light laser. Our results show that the topological phase of the system can be switched from one topological phase to another one via varying the light intensity. The intrinsic DMI plays a crucial role in the occurrence of photoinduced topological phase transition. It is shown that the sign reversal of the thermal hall conductivity is an important indicator on photoinduced topological phase transitions in the Kitaev–Heisenberg honeycomb ferromagnet.

## 1. Introduction

In the past few decades, the two-dimensional quantum magnet has been regarded as a good platform to realize magnetic analogues of those topological phases undiscovered in the electronic materials[1-16]. The magnon (i.e., quantized spin wave) is a quasi-particle for low-energy collective mode in quantum magnet[17,18]. Until now, more and more researchers have paid attention to topological properties of magnons in insulating quantum magnet, and the emergences of chiral magnon edge state and bulk Chern number have been proved[19-22].

[*] Corresponding author.
E-mail addresses: bingtangphy@jsu.edu.cn

Topological magnon insulators are the bosonic version of topological insulators in the electronic system[22]. They possess some interesting topological properties, e. g., supporting the nontrivial energy band structure and topologically protected edge state[19-21]. Experimentally, Onose *et al.* have first identified the magnon thermal Hall effect in one insulating ferromagnet $Lu_2V_2O_7$ with a pyrochlore structure[23]. They have found that the transverse heat current can emerge when a longitudinal temperature gradient is introduced. The magnon Hall effect is caused by the presence of edge magnon currents in the two-dimensional (2D) quantum magnetic systems[24,25]. Zhang *et al.* [19] have shown that the magnon edge current is resulted from the nontrivial topology of magnon bands, and first realized a topological magnon insulator on the pyrochlore lattice. Up to now, topological magnon insulators have also been theoretically realized in other 2D Heisenberg ferromagnets, such as the kagome[21], honeycomb[26], and Lieb ferromagnets [27].

Physically, uncharged magnons in the quantum ferromagnet can be viewed as a magnetic dipole moment hopping on the lattice[28,29]. The magnetic dipole moving under an applied electric field can accumulate a Aharonov-Casher phase [30]. This physical phenomenon is called the Aharonov-Casher effect, which is analogous to Aharonov-Bohm effect in electronic systems. Experimentally, Zhang *et al.* [31] have observed the Aharonov-Casher effect for the magnons of the yttrium iron garnet under an oscillating applied electric field, which can give rise to some interesting results. Owerre has studied Floquet topological magnons in 2D laser-irradiated Heisenberg ferromagnets[32,33]. For the Heisenberg ferromagnet, a circular-polarized laser can induce a tunable Dzyaloshinskii-Moriya interaction (DMI)[34,35], which may cause a photoinduced topological phase transition. Very recently, we have found that the photoinduced topological phase transition can occur in a 2D checkerboard Heisenberg ferromagnet, which is irradiated by the circular-polarized laser[36].

In this letter, we will focus on Floquet topological magnons in a 2D laser-irradiated Kitaev-Heisenberg honeycomb ferromagnet with the DMI. In the realistic 2D magnetic material, Kitaev interaction is usually accompanied by DMI. Hence, it is important to understand the role of DMI in the 2D Kitaev-Heisenberg magnet. Fortunately, the interplay mechanisms of these two interactions for magnonic properties of Heisenberg-Kitaev ferromagnets have been well clarified[37,38]. Here, our aim is to utilize a circular-polarized laser to introduce some tunable parameters in the effective Hamiltonian of the system, which may lead to the topological phase transition of Floquet magnons. According to the result from Ref. [39], we have realized that photoinduced topological phase transitions do not occur in the laser-irradiated Kitaev-Heisenberg honeycomb ferromagnet without the DMI. Hence, we will study the effect of the intrinsic DMI on the photoinduced topological phase transitions in the 2D Heisenberg-Kitaev ferromagnet. More details will be shown in the remainder of the paper.

## 2. Model

In this research, we consider an extended ferromagnetic Kitaev-Heisenberg model on the honeycomb lattice, as displayed in Fig. 1. The corresponding spin Hamiltonian is given by

$$\mathcal{H} = J\sum_{\langle i,j \rangle} \vec{S}_i \cdot \vec{S}_j + 2K \sum_{\langle i,j \rangle \gamma} S_i^\gamma S_j^\gamma + \sum_{\langle\langle i,j \rangle\rangle} \vec{D}_{ij} \cdot \left( \vec{S}_i \times \vec{S}_j \right). \qquad (1)$$

Here, the first and second terms denote the first-nearest-neighbours Heisenberg ferromagnetic exchange interaction ( $J<0$ ) and the bond-dependent Kitaev interaction, respectively. The index $\gamma \in \{X,Y,Z\}$ stands for the bond type as shown in Fig. 1. These two interactions can be parameterized as $J = A\cos\theta$ and $K = A\sin\theta$. When $0.85\pi < \theta < 3\pi/2$, the system exhibits the ferromagnetic phase[8]. The third term corresponds to the second-nearest-neighbours antisymmetric DMI. $\vec{D}_{ij} = Dv_{ij}\vec{e}_z$ is the DMI vector, where $\vec{e}_z$ is the unit vector along the cubic [111] direction and the coefficient $v_{ij} = -v_{ji} = \pm 1$ depends on the orientation of the two second-nearest-neighbours spins.

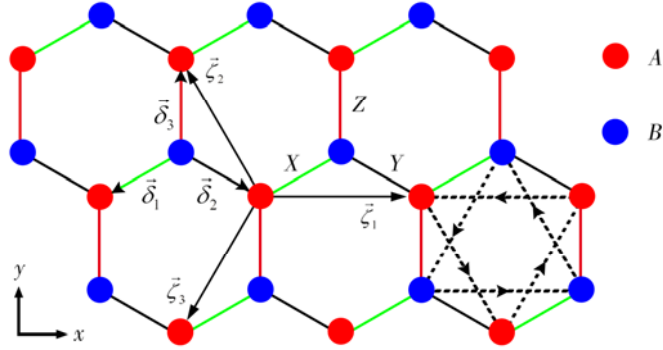

**Fig. 1.** (Color online) Schematic of the honeycomb lattice of the Kitaev-Heisenberg, which is made up of two different sublattices A (red circle) and B (Blue circle). The Kitaev bonds $X$ (green), $Y$ (black), $Z$ (red) are indicated with thick colored lines. For convenience, we need to introduce a new orthonormal basis $(\vec{e}_x, \vec{e}_y, \vec{e}_z)$, where $\vec{e}_z$ is parallel to the cubic [111] direction, i.e., perpendicular to the two-dimensional honeycomb lattice plane [40]. $\vec{\delta}_n, \vec{\varsigma}_n$ $(n=1,2,3)$ are the three NN and NNN vectors, respectively.

### 3. Theory of laser-Irradiated Kitaev magnets

*3.1. The Aharanov Casher phase*

Physically, when the quantum magnet is illuminated by one intensive laser light, the spin is able to couple to the light field via diverse processes [33,41-43]. As is well-known, magnons are quantized spin waves, which are in fact oscillations of those spin magnetic dipole moments. If the spin magnetic dipole moment is quantized along the $\vec{e}_z$ direction, then it can be expressed as $\vec{\mu}_S = g\mu_B \vec{e}_z$. Under the intense laser field with one oscillating (time-periodic) electric field $\vec{E}(\tau)$, the hopping spin

magnetic dipole moment $\vec{\mu}_S$ shall scrape up one time-dependent AC phase [30], which is given by

$$\phi_{ij}(\tau) = \frac{g\mu_B}{\hbar c^2} \int_{\vec{r}_i}^{\vec{r}_j} \vec{\Xi}(\tau) \cdot d\vec{l}. \tag{2}$$

Here, $\vec{\Xi}(\tau) = \vec{E}(\tau) \times \vec{e}_z$ with $\vec{E}(\tau) = -\partial_\tau \vec{A}(\tau)$, $\vec{A}(\tau)$ denotes the electromagnetic vector potential for the external laser field.

In the present work, we consider that the laser beam propagates along the $\vec{e}_z$ direction, where the oscillating electric field has the following form

$$\vec{\Xi}(\tau) = E_0 \left[ \sin(\omega\tau), \sin(\omega\tau + \Theta), 0 \right]. \tag{3}$$

Here, $E_0$ stands for the amplitude of the electric field, $\omega$ represents the circular frequency of the light wave, and $\Theta$ is the phase difference. In the current work, we consider the circularly-polarized light, i.e., $\Theta = \pi/2$.

In the new basis, one can obtain a time-dependent version of the spin Hamiltonian, which reads

$$\mathcal{H}(\tau) = \left(J + \frac{2K}{3}\right) \sum_{\langle i,j \rangle} \left[ S_i^z S_j^z + \frac{1}{2}\left( S_i^+ S_j^- e^{i\phi_{ij}(\tau)} + H.c. \right) \right]$$
$$+ \frac{2K}{3} \sum_{\langle i,j \rangle_\gamma} \frac{1}{2} \left[ e^{i\varphi_\gamma} S_i^+ S_j^+ e^{i\phi_{ij}(\tau)} + H.c. \right]$$
$$+ i\frac{1}{3} D \sum_{\langle\langle i,j \rangle\rangle} v_{ij} \frac{1}{2} \left[ S_i^- S_j^+ e^{-i\phi_{ij}(\tau)} - H.c. \right],$$

(4)

where $S_i^\pm = S_i^x \pm i S_i^y$ stand for the spin raising and lowering operators, and $\varphi_\gamma$ originates from the rotation of the bond directions. Note that $\varphi_\gamma = \frac{2\pi}{3}, \frac{4\pi}{3}, 0$ correspond to the $X$, $Y$, $Z$ three bond directions, respectively. We focus on the linear spin-wave approximation, which is reasonable in the large spin value limit and the low-temperature regime. This can be implemented via recasting the spin operators in the time-dependent Hamiltonian in terms of the following linearized Holstein Primakoff (HP) transformation [44]

$$S_i^+ = \sqrt{2S} a_i, \quad S_i^- = \sqrt{2S} a_i^+, \quad S_i^z = S - a_i^+ a_i. \tag{5}$$

Here, $a^+(a)$ is the magnon creation (annihilation) operator. The resulting linear bosonic Hamiltonian has time periodicity, i.e., $\mathcal{H}(\tau+T) = \mathcal{H}(\tau)$, where $T = \frac{2\pi}{\omega}$ corresponds to the period of the laser field.

*3.2. Floquet-Bloch Hamiltonian for the laser driven Kitaev-Heisenberg model*

For the sake of investigating the periodically driven Kitaev-Heisenberg honeycomb ferromagnet in Eq. (4), we will apply the Floquet theory[45] to transform the present time-dependent spin model to one static time-independent effective spin model, which is describe by the Floquet Hamiltonian. Physically, this static effective Hamiltonian $\mathcal{H}_{eff}$ can be expressed in terms of $\omega^{-1}$, namely, $\mathcal{H}_{eff} = \sum_{\eta \geq 0} \omega^{-\eta} \mathcal{H}^\eta$, where $\eta$ is an integer. When the circular frequency $\omega$ of the laser is much larger than the magnon frequency bandwidth $\Delta$, i.e., $\omega \gg \Delta$, this way is applicable. In the current work, we pay attention to the off-resonant regime, thus it suffices to take into account the zeroth order of the static Floquet Hamiltonian [33]. By making use of the discrete Fourier component of the time-dependent Hamiltonian, we can obtain a zeroth order effective Hamiltonian $\mathcal{H}_{eff} = \mathcal{H}^0$, where $\mathcal{H}^\eta = \frac{1}{T} \int_0^T d\tau e^{-i\eta\omega\tau} \mathcal{H}(\tau)$ and $\eta$ is an integer. In the momentum space, the effective Hamiltonian can be written as $\mathcal{H} = \sum_{\vec{k}} \psi_{\vec{k}}^+ H(\vec{k}) \psi_{\vec{k}}$, where the basis

$\psi_{\vec{k}} = \left(a_{\vec{k}}, b_{\vec{k}}, a^+_{-\vec{k}}, b^+_{-\vec{k}}\right)^T$. The Hamiltonian matrix $H(\vec{k})$ takes the following form

$$H(\vec{k}) = \frac{S}{2}\begin{pmatrix} A^0(\vec{k}) & B^0(\vec{k}) \\ \left[B^0(\vec{k})\right]^+ & \left[A^0(-\vec{k})\right]^T \end{pmatrix} \tag{6}$$

with

$$A^0(\vec{k}) = \begin{pmatrix} \rho_0(\vec{k}) & \rho_1(\vec{k}) \\ \rho_1(\vec{k})^* & \rho_0'(\vec{k}) \end{pmatrix}, \tag{7}$$

$$B^0(\vec{k}) = \begin{pmatrix} 0 & \rho_2(\vec{k}) \\ \rho_2(-\vec{k}) & 0 \end{pmatrix}, \tag{8}$$

where

$$\rho_0(\vec{k}) = \left\{-3J - 2K - \frac{2}{3}D\left[J_0\left(\sqrt{3}\varepsilon_0\right)\sin\left(\vec{k}\cdot\vec{\varsigma}_1\right) + J_0\left(\sqrt{3}\varepsilon_0\right)\sin\left(\vec{k}\cdot\vec{\varsigma}_2\right) + J_0\left(\sqrt{3}\varepsilon_0\right)\sin\left(\vec{k}\cdot\vec{\varsigma}_3\right)\right]\right\},$$

$$\rho_0'(\vec{k}) = \left\{-3J - 2K + \frac{2}{3}D\left[J_0\left(\sqrt{3}\varepsilon_0\right)\sin\left(\vec{k}\cdot\vec{\varsigma}_1\right) + J_0\left(\sqrt{3}\varepsilon_0\right)\sin\left(\vec{k}\cdot\vec{\varsigma}_2\right) + J_0\left(\sqrt{3}\varepsilon_0\right)\sin\left(\vec{k}\cdot\vec{\varsigma}_3\right)\right]\right\},$$

$$\rho_1(\vec{k}) = \left(J + \frac{2K}{3}\right)\left[J_0(\varepsilon_0)e^{i\vec{k}\cdot\vec{a}_1} + J_0(\varepsilon_0)e^{i\vec{k}\cdot\vec{a}_2} + J_0(\varepsilon_0)\right],$$

$$\rho_2(\vec{k}) = \frac{2K}{3}\left[J_0(\varepsilon_0)e^{i\vec{k}\cdot\vec{a}_1 + i\frac{2\pi}{3}} + J_0(\varepsilon_0)e^{i\vec{k}\cdot\vec{a}_2 - i\frac{2\pi}{3}} + J_0(\varepsilon_0)\right]. \tag{9}$$

Here, $J_\eta(x)$ is known as Bessel function of the order $\eta \in Z$. In fact, the light intensity of the laser can be characterized via a dimensionless quantity $\varepsilon_0 = g\mu_B E_0 / \hbar c^2$. It is noted that the static effective magnon Hamiltonian in Eq. (6) is analogous to the Bogoliubov-de-Gennes Hamiltonian in the context the superconductivity. Theoretically, this Hamiltonian matrix can be diagonalized via making use of one paraunitary Bogoliubov transformation.

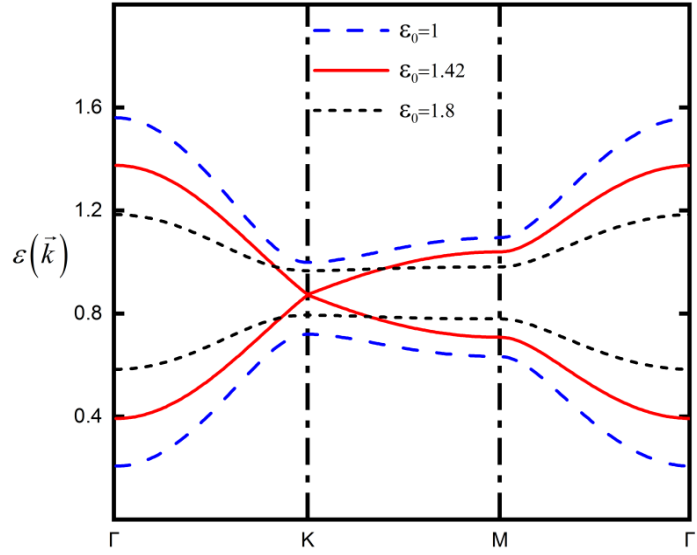

**Fig. 2.** (Color online) The Floquet topological magnon of three different the light intensities linear dispersion relation along the path $\Gamma-K-M-\Gamma$. The other parameters are chosen as $\theta=\dfrac{5\pi}{4}$, $A=1$, $D=0.7$, and $S=\dfrac{1}{2}$.

In Fig. 2, we display the Floquet topological magnon (linear) dispersion relation of the Kitaev-Heisenberg honeycomb ferromagnetic lattice. Our results indicate that the band gap at the Dirac point $K$ decreases and the spectrum becomes gapless at $\varepsilon_0=\varepsilon_c\approx 1.42$ as the light intensity increases. If one further the increases light intensity, the band gap reopens. Physically, when the gap at the K-point opens, the present system can be viewed as a well-defined honeycomb topological magnon insulator. Here, we clearly see that the laser can derive a gap-closing phenomenon, which means a continuous topological phase transition may occur with the increase of the light intensity.

*3.3. Laser-induced topological transitions*

In those topological magnon systems, the Berry curvature is the fundamental of a great many observables. Physically, one nontrivial band topology can emerge only when the magnetic system reveals a band gap between two different magnon bands, and a non-vanishing Chern number implies the emergence of magnon edge state. In two 2D magnetic system, the Berry curvature of the n-th magnon band has the following form

$$\Omega_n(\vec{k}) = -2\,\mathrm{Im}\sum_{m\neq n} \frac{\left[\langle T_n(\vec{k})|\hat{v}_x|T_m(\vec{k})\rangle\langle T_m(\vec{k})|\hat{v}_y|T\psi_n(\vec{k})\rangle\right]}{\left[\varepsilon_n(\vec{k})-\varepsilon_m(\vec{k})\right]^2}. \qquad (10)$$

Here, $T(\vec{k})$ is a paraunitary matrix, $T_n(\vec{k})$ corresponds to its n-th component, $\hat{v}_i = \partial \mathcal{H}_B(\vec{k})/\partial k_i\,(i=x,y)$ is called the velocity operator, and $\varepsilon_n(\vec{k})$ is the n-th magno energy band. Moreover, the corresponding bosonic Bogoliubov Hamiltonian is given by $H_B(\vec{k}) = \tau_3 H(\vec{k})$ with $\tau_3 = [(I_{N\times N}, 0), (0, -I_{N\times N})]$. In order to address the topological property of Floquet magnons, we need to compute the associated Chern number, which is defined as

$$C_n = \frac{1}{2\pi}\int_{BZ} d^2k\,\Omega_n^z(\vec{k}). \qquad (11)$$

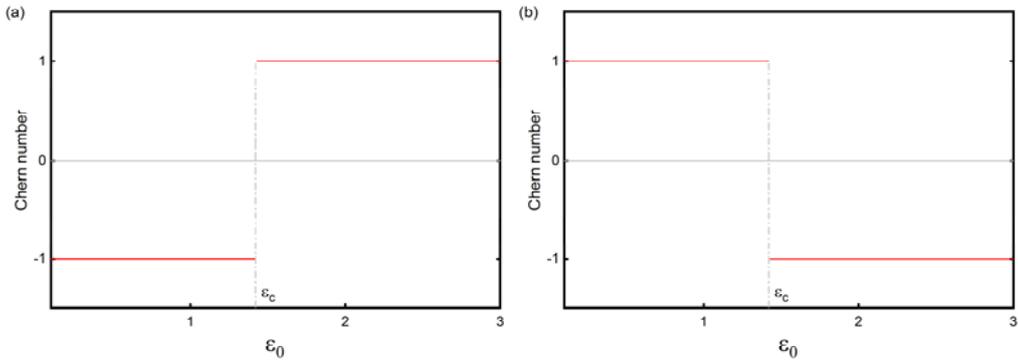

**Fig. 3.** (Color online) The Chern numbers of the Floquet topological magnon bands

versus#the light intensity in the honeycomb Kitaev-Heisenberg ferromagnet: (a) the optical "up" band, (b) the acoustic "down" band. Here, the critical value $\varepsilon_c$ is approximately equal to 1.42. The corresponding parameters are set to $\theta = \frac{5\pi}{4}$, $D = 0.7$, $A = 1$, and $S = \frac{1}{2}$.

In Fig. 3, we display the light intensity $\varepsilon_0$ dependence of the Chern numbers for two Floquet topological magnon bands in the honeycomb Kitaev-Heisenberg ferromagnet. It is clearly seen that the present honeycomb ferromagnet switches from one Floquet topological magnon insulator having Chern numbers $(C_u, C_d) = (-1, 1)$ to another one having Chern numbers $(C_u, C_d) = (1, -1)$ by increasing the light intensity $\varepsilon_0$. Under the current parameter settings, there exist a topological phase transition at the critical light intensity $\varepsilon_0 = \varepsilon_c \approx 1.42$, which corresponds to the band gap closes.

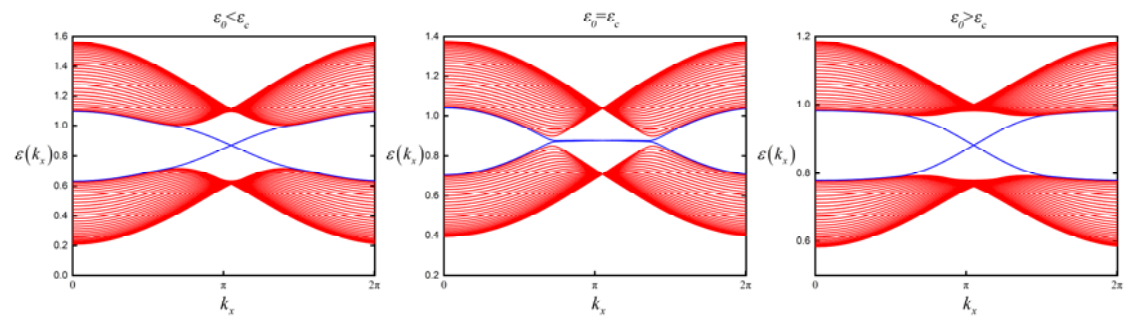

**Fig. 4.** (Color online) Floquet magnon energy dispersions for bulk and edge states in the laser-irradiated honeycomb Kitaev-Heisenberg ferromagnet possessing a zig-zag edge at different light intensities. The relevant parameters are the same as in Fig. 2.

As is well-known, one of the natures of 2D topological magnon systems is the presence of the chiral magnon edge state [46,47]. In the 2D insulating topological ferromagnet, the chiral edge state plays an important role in the spin transport, which is caused by the topological characteristic of the continuous magnon bluk bands. By applying numerical diagonalization technique, one can compute the Floquet magnon energy spectrums of a zigzag honeycomb Kitaev-Heisenberg ferromagnet for various light intensities, as presented in Fig. 4. In the case of $\varepsilon_0 < \varepsilon_c$ a magnon band gap occurs at the Dirac point and the gap closes around the first null point of the Bessel function $\varepsilon_0 = \varepsilon_c$ and it reopens when $\varepsilon_0 > \varepsilon_c$. In Fig. 4, it is clearly seen that topologically protected chiral magnon edge states can exist between two magnon bulk bands for $\varepsilon_0 < \varepsilon_c$ or $\varepsilon_0 > \varepsilon_c$. It is noted that there are no chiral magnon edge states for $\varepsilon_0 = \varepsilon_c$.

*3.4. Laser-induced magnon thermal Hall effects*

The magnon thermal Hall effect is viewed as one significant consequence of the 2D topological magnon insulator. Physically, if one longitudinal temperature gradient emerges, then the presence of the Floquet magnon edge states can lead to one transverse heat current. This Floquet magnon edge flux shall produce a Floquet magnon thermal Hall effect. In principle, the Floquet magnon thermal Hall effect can be characterized via a thermal Hall conductivity, which is related to the Berry curvature for the Floquet topological magnon eigenstates of the system[33]. Here, we mainly pay attention to the regime, in which Bose occupation function approaches the thermal equilibrium. In this regime, the Floquet thermal Hall conductivity has the following form[24,33,48]

$$\kappa_{xy} = -\frac{k_B^2 T}{(2\pi)^2 \hbar} \sum_{\alpha=m,n} \int d^2k\, c_2(n_\alpha) \Omega_\alpha(\vec{k}), \qquad (12)$$

where $n_\alpha = f(\varepsilon_\alpha(\vec{k})) = \left[e^{\beta \varepsilon_\alpha(\vec{k})} - 1\right]^{-1}$ is famous Bose-Einstein distribution function, $\beta = 1/k_B T$, $c_2(w) = (1+w)\left(\ln\frac{1+w}{w}\right)^2 - (\ln w)^2 - 2\mathrm{Li}_2(-w)$. Here, it is noted that $\mathrm{Li}_2(x)$ corresponds to the dilogarithm function.

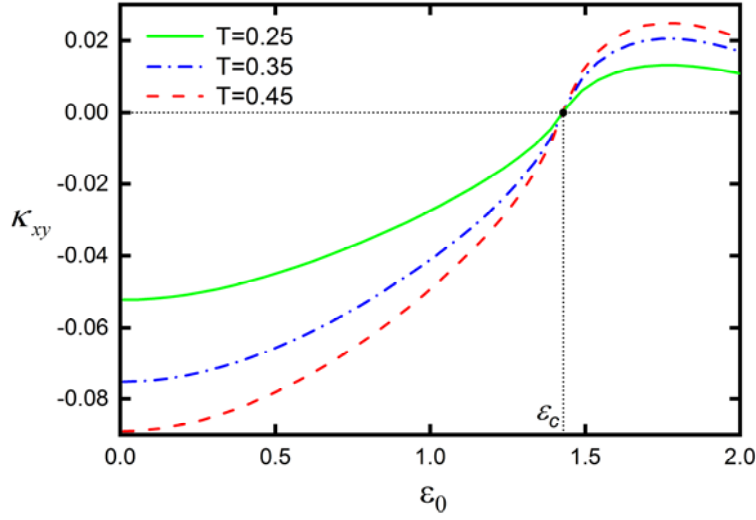

**Fig. 5.** (Color online) The thermal Hall conductivity vs the light intensity for various values of the temperature. The other parameters are the same as in Fig. 2.

Physically, the magnon thermal Hall conductivity $\kappa_{xy}$ changes sign when laser-irradiated Kitaev-Heisenberg ferromagnets undergo one topological phase transition. In Fig. 5, we show the thermal Hall conductivity $\kappa_{xy}$ versus the light intensity $\varepsilon_0$ for various values of temperature $T$. We can clearly see that the thermal Hall conductivity $\kappa_{xy}$ changes continuously with the light intensity $\varepsilon_0$ and goes from being negative for small $\varepsilon_0$ ($\varepsilon_0 < \varepsilon_c$) to being positive for big $\varepsilon_0$ ($\varepsilon_0 > \varepsilon_c$). We note that the thermal hall conductivity and the Chern number of the upper band have the same signs with the increase of the light intensity $\varepsilon_0$. That is to say, the sign reversal of $\kappa_{xy}$ is a significant indicator to identify the topological phase transition of Floquet magnons in the laser-irradiated Kitaev-Heisenberg honeycomb ferromagnet.

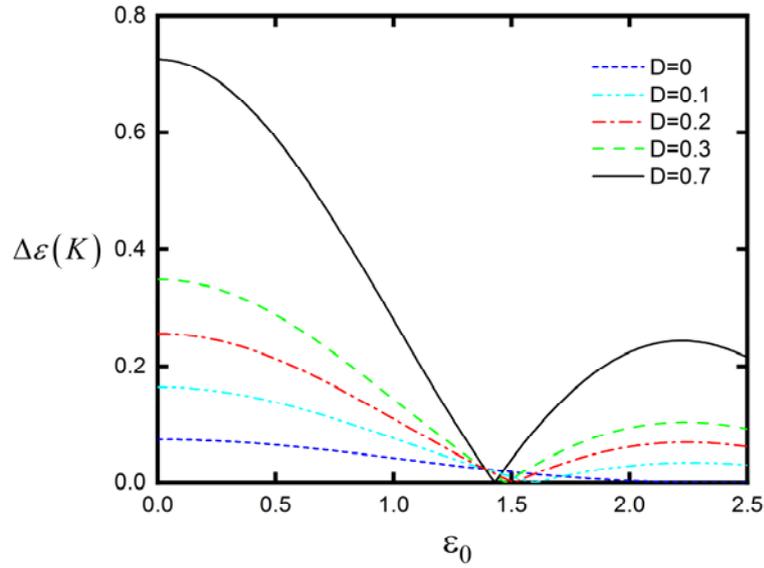

**Fig. 6.** (Color online) The gaps at the Dirac point as a function of the light intensity $\varepsilon_0$ for various values of $D$. The corresponding parameters are set to $\theta = \frac{5\pi}{4}$, $A = 1$, and $S = \frac{1}{2}$.

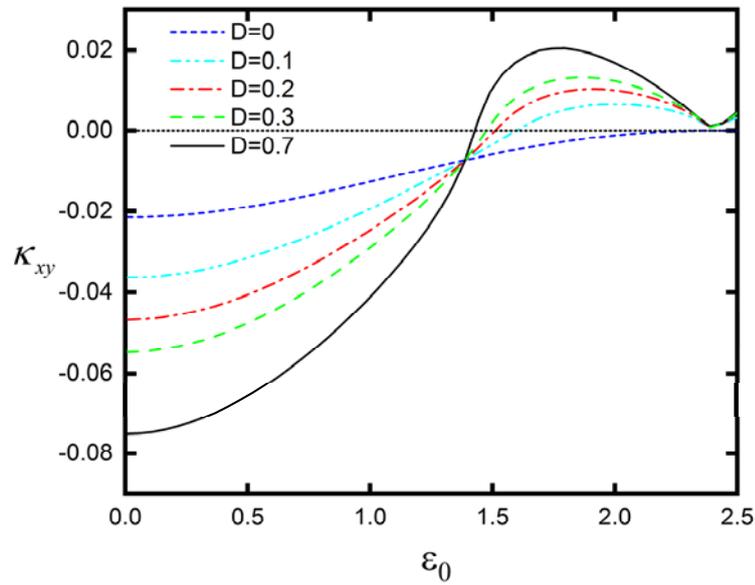

**Fig. 7.** (Color online) The thermal Hall conductivity vs the light intensity for various values of $D$. The temperature is fixed as $T = 0.35$, and the other parameters are the same as in Fig. 6.

## 4. Discussion and summary

Now, we discuss the effect of the DMI on the photoinduced topological phase transitions in the Kitaev-Heisenberg honeycomb ferromagnet. In order to understand the influence of the DMI strength on sizes of the gap, we plot the gaps at the Dirac point as a function of the light intensity for different values of $D$, as displayed in Fig. 6. When the light intensity is fixed, the band gap width increases with the increase of $D$. Moreover, we note that the critical light intensity corresponding to the gap-closing decrease as increasing the value of $D$. As mentioned previously, the sign reversal of thermal Hall conductivity is a significant indicator on the photoinduced topological phase transition. In Fig. 7, we show the dependence of the thermal Hall conductivity on the light intensity for various values of $D$. It is clearly seen that the sign of $\kappa_{xy}$ can change with the increase of $\varepsilon_0$ only when the DMI is introduced. As the value of $D$, the critical light intensity for the sign reversal of $\kappa_{xy}$ decreases. In the absence of the DMI, the magnon gap does not close and reopen when the m light intensity continuously increases. Thus, it is easy to understand the sign reversal of $\kappa_{xy}$ can not appear in the laser-irradiated Kitaev–Heisenberg honeycomb ferromagnet without DMI. Hence, the DMI plays a vital role in the photoinduced topological phase transitions in the present honeycomb ferromagnet.

To summarize, the topological phase transitions in a laser-irradiated Kitaev–Heisenberg honeycomb ferromagnet with DMI was theoretically investigated by means of the Floquet-Bloch theory. In the presence of a circular-polarized light, the system can possess two different topological phases, namely, $(C_u, C_d) = (-1, 1)$ and $(C_u, C_d) = (1, -1)$. The topological phase of the present laser-irradiated ferromagnetic system can be controlled by the light intensity. This continuous photoinduced topological phase transition is along with the gap-closing phenomenon.

It was shown that and the Chern number for the upper band and the thermal hall conductivity have the same signs. Thus, the thermal Hall conductivity directly can reflect the topological phase of the present ferromagnetic system. In future experiments, the sign reversal of the thermal hall conductivity can be identified as a significant indicator to observe the photoinduced topological phase transition in the 2D Kitaev-Heisenberg honeycomb ferromagnet.

**Acknowledgments**

We want to thank Dr. Kangkang Li for a lot of help. This work was supported by the National Natural Science Foundation of China under Grant No. 12064011 and the Graduate Research Innovation Foundation of Jishou University under Grant No. Jdy21030.